\documentclass[conference]{IEEEtran}

\usepackage{multirow}
\ifCLASSINFOpdf
   \usepackage[pdftex]{graphicx}
   
   \graphicspath{{./pdf/}{./jpeg/}}
   \DeclareGraphicsExtensions{.pdf,.jpeg,.png}
\else
   \usepackage[dvips]{graphicx}
   \graphicspath{{./eps/}}
   \DeclareGraphicsExtensions{.eps}
\fi
\usepackage{mathtools}
\usepackage{color}
\usepackage{algorithm}
\usepackage{algpseudocode}

\usepackage{times}
\usepackage{epsfig}
\usepackage{graphicx}
\usepackage{amsmath}
\usepackage{amssymb}
\usepackage{romannum}
\usepackage{color}
\usepackage{caption}
\usepackage{subcaption}

\usepackage{amsmath}
\usepackage{multirow}

\usepackage{algorithm,algpseudocode}

\usepackage{url}
\begin{document}
\title{Enhancing LLM Fine-tuning for Text-to-SQLs by SQL Quality Measurement}

\author{\authorblockN{Shouvon Sarker, Xishuang Dong, Xiangfang Li, Lijun Qian}
\authorblockA{CREDIT Center and Department of Electrical and Computer Engineering \\
Prairie View A\&M University\\
Prairie View, TX 77446, USA \\
Email: ssarker3@pvamu.edu, xidong@pvamu.edu, xili@pvamu.edu, liqian@pvamu.edu }
}


%


\maketitle

\begin{abstract}
%\boldmath

Text-to-SQLs enables non-expert users to effortlessly retrieve desired information from relational databases using natural language queries. While recent advancements, particularly with Large Language Models (LLMs) like GPT and T5, have shown impressive performance on large-scale benchmarks such as BIRD, current state-of-the-art (SOTA) LLM-based Text-to-SQLs models often require significant efforts to develop auxiliary tools like SQL classifiers to achieve high performance. This paper proposed a novel approach that only needs SQL Quality Measurement to enhance LLMs-based Text-to-SQLs  performance. It establishes a SQL quality evaluation mechanism to assess the generated SQL queries against predefined criteria and actual database responses. This feedback loop enables continuous learning and refinement of model outputs based on both syntactic correctness and semantic accuracy. The proposed method undergoes comprehensive validation on the BIRD benchmark, assessing Execution Accuracy (EX) and Valid Efficiency Score (VES) across various Text-to-SQLs difficulty levels. Experimental results reveal competitive performance in both EX and VES compared to SOTA models like GPT4 and T5. 
\end{abstract}
\begin{IEEEkeywords} Text-to-SQLs; Large Language Models (LLMs); SQL Quality Measurement; Database Information Retrieval  \end {IEEEkeywords}

% \fi
\IEEEpeerreviewmaketitle

%review
\section{Introduction}
\label{sec1}

%\section{Introduction}
% Text-to-SQLs

\begin{figure*} [ht]
 	\centering
	\includegraphics[width=1.\linewidth]{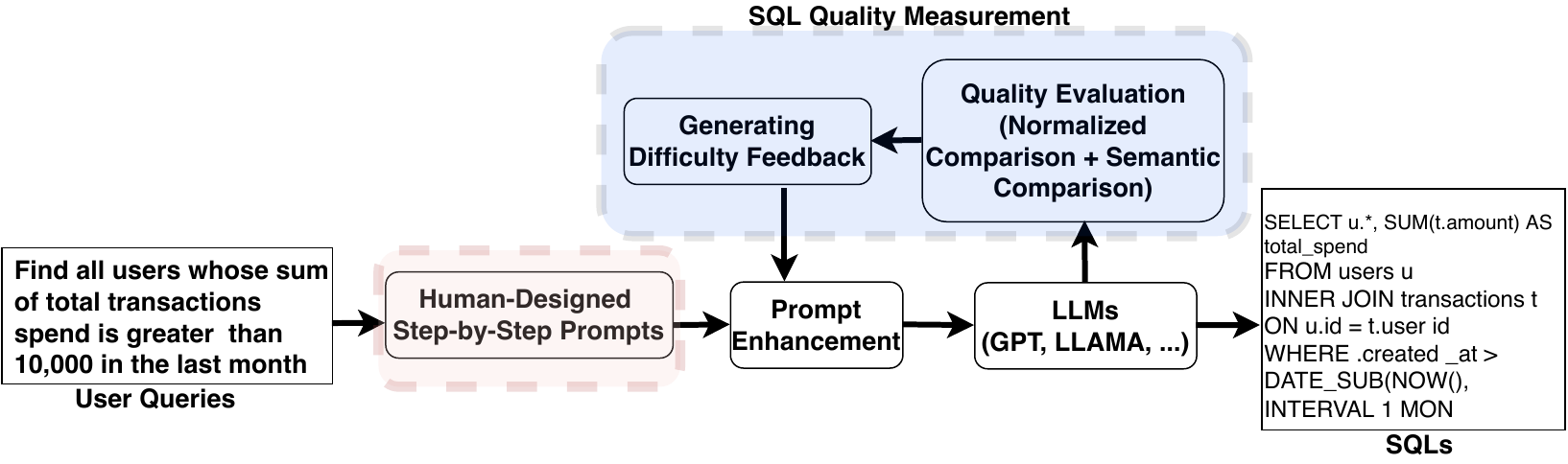}
	\caption{Flow of Enhancing LLM Fine-tuning for Text-to-SQLs by SQL Quality Measurement.}
	\label{Fig_framework}
\end{figure*}

Text-to-SQLs~\cite{ qin2022survey, kumar2022deep, zhang2024benchmarking, li2024can} aims to convert users' queries into executable and structured SQL statements, where these queries are presented in natural languages. This technique enables non-expert users, particularly those are not familiar with the SQL language, to seamless access to knowledge stored in rational databases for applications in various areas such as finance and medicine. Nonetheless, it poses greater challenges compared to general semantic parsing tasks~\cite{kumar2022deep, li2024can}. First, even simple user queries may involve complex combinations of multiple tables and filtering requirements, necessitating sophisticated context understanding techniques to fulfill query needs. Secondly, there remains a significant gap between existing Text-to-SQL methods and real-world applications in efficiently and effectively generating high-quality SQLs, especially when dealing with large-scale databases.

Deep learning technologies have greatly advanced Text-to-SQLs methods through developing both non-seq2seq methods and seq2seq methods classified in terms of  model architectures~\cite{kumar2022deep, qin2022survey}. For non-seq2seq methods~\cite{cai2021sadga, choi2021ryansql, hui2022s2sql},  they typically involve two stages: 1) it employ encoder models with attention mechanisms like BERT~\cite{devlin2018bert} to learn high-quality representations of user queries; 2) it constructs sketch-based or grammar-based systems to generate SQL statements. On the other hand,  seq2seq methods~\cite{lin2020bridging, qi2022rasat, li2023graphix}  treat Text-to-SQLs as a machine translation task, directly translating user queries into SQL statements in an end-to-end manner. This approach has achieved competitive performance through fine-tuning with minimal effort. However, a reliable Text-to-SQLs parser has yet to be fully implemented.

The emergence of Large Language Models (LLMs) has recently facilitated Text-to-SQL methods by building an interface  between non-expert users and relational databases, leveraging the robust linguistic and coding capabilities of LLMs. This has sparked a new trend of \textit{LLM-based Text-to-SQL approaches}~\cite{zhang2024benchmarking}. Specifically, the in-context learning and domain generalization abilities of LLMs have significantly improved Text-to-SQL performance. For example, DIN-SQL decomposes the Text-to-SQL task into smaller sub-tasks to enhance performance on datasets like Spider~\cite{pourreza2024din}. DAIL-SQL explores the selection and organization of helpful examples in prompts in few-shot scenarios through supervised fine-tuning and a systematic study of in-context learning~\cite{gao2023text}. ACT-SQL introduces a Chain-of-Thought (CoT)~\cite{wei2022chain} prompt to enhance reasoning abilities in SQL generation, extending to multi-turn Text-to-SQL tasks~\cite{zhang2023act}. Furthermore, to address database schema with few rows of values, the Big bench for large-scale Database (BIRD) is proposed to bridge the gap between academic study and real-world applications~\cite{li2024can}. Text-to-SQL methods applied on BIRD have garnered significant attention~\cite{li2024can, pourreza2024dts, wang2023mac} , achieving high performance through additional efforts such as developing SQL classifiers~\cite{li2024can}, fine-tuning  LLMs~\cite{pourreza2024dts}, or facilitating multi-agent LLM collaborations~\cite{wang2023mac}. Although these research efforts have advanced LLM-based Text-to-SQL systems, they often require complex data preprocessing or data augmentation techniques.

In this study, we aim to enhance LLM-based Text-to-SQL by using straightforward feedback mechanisms and human-designed step-by-step prompts (HDSP).  The method requires only LLMs reasoning and low-cost SQL quality assessment. The flow of the proposed method is illustrated in Figure~\ref{Fig_framework}. First, in the prompt engineering block, we design step-by-step prompts to LLMs for enhancing the Text-to-SQL performance. Next, we implement a feedback mechanism through SQL quality measurement. This mechanism compares generated SQL statements against expected outcomes, providing direct feedback on SQL difficulties to automatically refine the prompts. We systematically validate our approach on BIRD datasets, assessing Execution Accuracy (EX) and Valid Efficiency Score (VES)~\cite{li2024can}. Experimental results demonstrate significant improvements in both metrics compared to state-of-the-art models like GPT4~\cite{achiam2023gpt}.

%\subsection{Contributions}
In summary, our contributions are as follows:

\begin{itemize}
\item We design step-by-step prompts that provide detailed instructions to LLMs to complete the Text-to-SQLs task effectively.
\item We implement SQL statement quality feedback by comparing generated SQL statements against expected outcomes, both normally and semantically, to inform automatic prompt refinement.
\item Extensive experiments are conducted to validate the proposed method, assessing accuracy and efficiency. Experimental results confirm that our approach not only automates prompt construction but also enhances prompts based on SQL statement quality, thereby improving Text-to-SQL performance.
\end{itemize}

\section{Task Formulation}
\label{sec6}
% task

LLMs-based Text-to-SQL involves utilizing LLMs to convert natural language user queries into executable SQL statements. Let $Q$ represent a natural language query and $D$ denote the database schema, where $D = (T , C)$ with $T = \{t_{1}, ... , t_{m}\}$ representing multiple tables and $C = \{c_{1}, ... , c_{n}\}$ representing columns. The objective is to generate a SQL statement $\hat{Y}$ that is executable on $D$ to retrieve the information required by $Q$. Formally, given a prompt template $P(Q, D)$ and external knowledge evidence $K$, the generation process of the SQL statement $\hat{Y}$ by a large language model (LLM) $M$ can be defined as a conditional probability distribution:

\begin{equation}
P_{M}(\hat{Y}|P(Q, D)) = \prod_{i = 1}^{|\hat{Y}|} P_{M}(\hat{Y}_{i}|\theta, P(Q, D), K, \hat{Y}_{1:i-1})
\end{equation}

It represents the likelihood of generating SQL statement $\hat{Y}$ given the natural language query $Q$, the database schema $D$, any external knowledge evidence $K$, and the prompt template $P$. The LLM generates each token, denoted by $\hat{Y}_{i}$ representing the $i$-th token of the SQL statement $\hat{Y}$. $|\hat{Y}|$ denotes the length of the SQL statement $\hat{Y}$, and $\theta$ represents the parameter of the LLM. Notably, $P_{M}(\hat{Y}|P(Q, D))$ is assessed in terms of syntactic correctness and semantic accuracy, defined as execution accuracy.

\section{Methodology}
\label{sec2}
% methodology

\begin{figure} [ht]
 	\centering
	\includegraphics[width=0.7\linewidth]{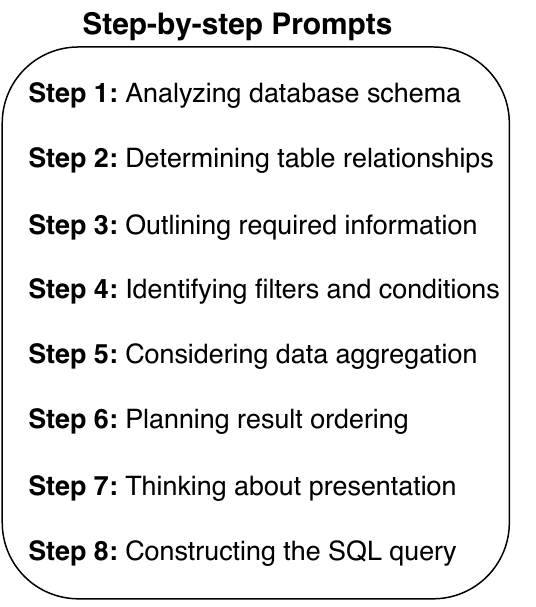}
	\caption{Human-designed step-by-step (HDSP) prompts for Text-to-SQLs via LLMs.}
	\label{Fig_prompt}
\end{figure}

As shown in Figure~\ref{Fig_framework}, the proposed method include two special components, namely, human-designed step-by-step prompts and SQL quality measurement.

\subsection{Step-by-step Prompts}

Figure~\ref{Fig_prompt} presents details of the step-by-step prompts for Text-to-SQLs via LLMs. It include eight steps as below.

\begin{enumerate}
\item LLMs are encouraged to begin by understanding the database schema. This involves identifying the available tables and their relevant columns for the user query, which is crucial for generating accurate SQL statements.

\item LLMs are prompted to identify how tables are interconnected by determining the types of joins needed to connect them, essential for SQL statements involving multiple tables.

\item LLMs consider the specific information to extract from each table, including which columns to select and any necessary calculations.

\item LLMs pinpoint any required filters or conditions to apply to the data, such as \texttt{WHERE} clauses, to refine the SQL statements according to the query's requirements.

\item LLMs assess whether the SQL statements involve aggregating data, potentially using \texttt{GROUP BY} or \texttt{HAVING} clauses to summarize or filter aggregated data.

\item LLMs determine how to order the SQL statements, which is crucial where the order of the SQL statements matters (e.g., the top 10 results).

\item LLMs consider the final presentation of the SQL statements, deciding whether to use subqueries for clarity or \texttt{WITH} clauses to simplify complex queries.

\item LLMs synthesize the information gathered in the previous steps into coherent SQL statements. 

\end{enumerate}

These step-by-step prompts are able to ensure that the SQL statements accurately reflect the natural language requirements of the user query, where LLMs can adjust their strategies, and gradually improve SQL generation capabilities. 

\subsection{SQL Quality Measurement}

As shown in Figure~\ref{Fig_framework}, it needs the quality of generated SQL to build a feedback of SQL difficulties to refine the prompts. Measuring the SQL quality includes two components: Normalized Comparison and Semantic Comparison.

\subsubsection{Normalized Comparison}

It quantifies the syntactic similarity between the SQL statement $\hat{Y}$ and the ground truth $Y$. Levenshtein distance ($L(\cdot)$) is employed as the metric for evaluating their similarity, defined as the minimum number of single-character edits required to change one string into another. Thus, the normalized comparison can be defined as:

\begin{equation}
\begin{split}
f_{nc}(\hat{Y}, Y) = 1 - \frac{L(f_{Norm}(\hat{Y}), f_{Norm}(Y))}{\max(\left| f_{Norm}(\hat{Y}) \right|, \left| f_{Norm}(Y) \right|)}
\end{split}
\end{equation}

where $f_{Norm}(\cdot)$ represents the normalization of an input.

\subsubsection{Semantic Comparison}

Semantic comparison  is defined as a binary indicator of whether the type of the answer $\hat{A}$ generated through running the generated SQL statement ($\hat{Y}$) is same to that of the answer ($A'$) generated by ChatGPT directly in terms of the user query, which is defined as:

\begin{equation}
f_{sc}(\hat{A}, A') =
\begin{cases} 
1 & \text{if } R(\hat{A}) = R(A')\\
0 & \text{otherwise}.
\end{cases}
\end{equation}

where $R(\cdot)$ is to check the type of the input.

In terms of the results of the normalized comparison, the feedback of SQL difficulties is defined by

\begin{equation}
f_{diff}(f_{nc}, N_{col}) = \begin{cases}
0 & if~~ f_{nc} < t_{nc}~~ and ~~ N_{col} \le 5\\
1 & if~~ f_{nc} < t_{nc}~~ and ~~ 5 < N_{col} < 10\\
2  & if~~ f_{nc} < t_{nc}~~ and ~~ N_{col} \ge 10 \\
\end{cases}
\end{equation}

where 0, 1, and 2 denote simple, moderate, and difficult SQL statements, $t_{nc}$ denotes a threshold of the normalized comparison, and $N_{col}$ refers to the number of columns involved in the SQL statement.  Furthermore, a specific feedback to implement the prompt enhancement shown in Figure~\ref{Fig_framework} is defined by

 \begin{equation}
f_{sf}(f_{nc}, f_{sc}, f_{diff}) = \begin{cases}
1 & if~~ f_{nc} \ge t_{nc}~~ and ~~ f_{sc} = 0\\
 & or ~~ f_{nc} < t_{nc}~~ and~~  f_{diff} = 1\\
 & or ~~ f_{nc} < t_{nc}~~ and~~  f_{diff} = 2\\
0  & otherwise \\
\end{cases}
\end{equation}

If the value of $f_{sf}$ is 1, the step-by-step prompts are extended with five steps below.

\begin{enumerate}

\item Complex Joins and Subqueries: LLMs are encouraged to plan the ordering of results and the combination of multiple datasets. This involves deciding on the appropriate \texttt{JOIN} types and conditions.

\item Data Transformation and Calculations: LLMs are prompted to identify any complex calculations or transformations needed for the data. This includes creating new columns based on existing data using SQL functions or handling \texttt{date} and \texttt{time} calculations.

\item Optimization: LLMs review the query for performance and optimization, considering indexing and query simplification to improve execution speed.

\item Security and Data Integrity: LLMs ensure that the query adheres to security practices and maintains data integrity, avoiding SQL injection risks by using parameterized queries or prepared statements.

\item Review and Test: Before finalizing, review the query thoroughly. Test it against different scenarios to ensure it covers all edge cases and produces expected results.

\end{enumerate}

\begin{figure} [ht]
 	\centering
	\includegraphics[width=0.65\linewidth]{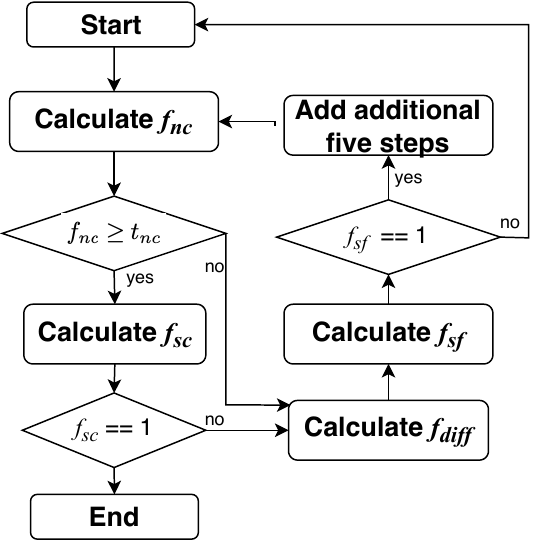}
	\caption{Flow of the proposed feedback mechanism.}
	\label{Fig_feedback}
\end{figure}

These five steps are inserted into the step-by-step prompts between step five and step six shown in Figure~\ref{Fig_prompt} to further enhance Text-to-SQLs. The flow of the proposed feedback mechanism is summarized as Figure~\ref{Fig_feedback}.

\section{Experiments}
\label{sec3}
\subsection{Dataset}

\begin{figure} [ht]
 	\centering
	\includegraphics[width=0.85\linewidth]{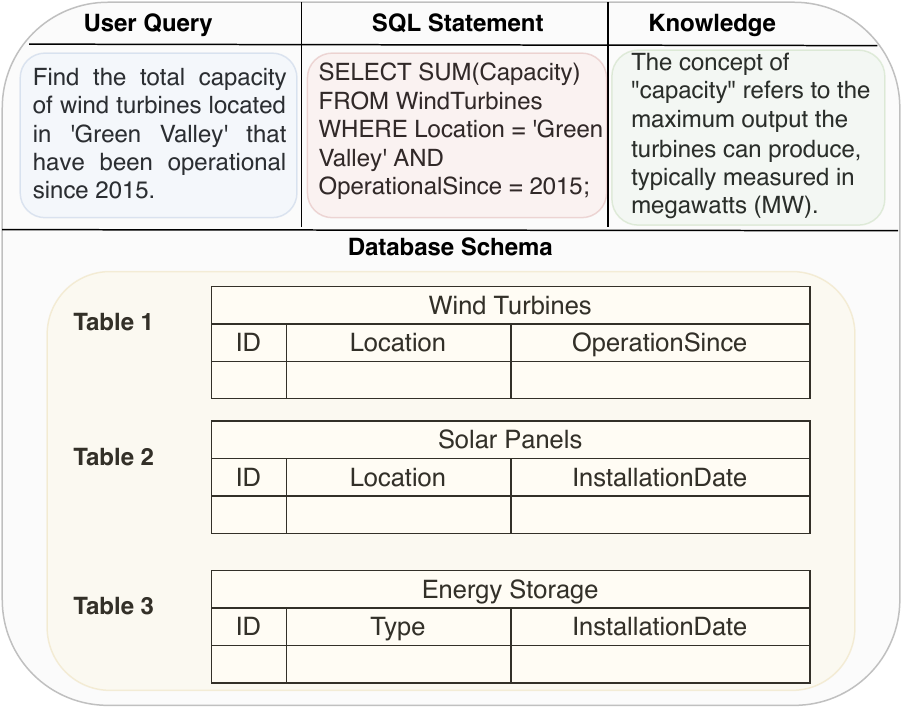}
	\caption{An example of data sample from BIRD benchmark.}
	\label{Fig_sample}
\end{figure}

BIRD (\textbf{BI}g Bench for La\textbf{R}ge-scale \textbf{D}atabase Grounded Text-to-SQL Evaluation)\footnote{https://bird-bench.github.io} is a pioneering, cross-domain dataset designed to examine the impact of extensive database contents on text-to-SQL parsing~\cite{li2024can}. BIRD includes over 12,751 unique question-SQL pairs and 95 large databases with a total size of 33.4 GB. It spans more than 37 professional domains, such as blockchain, hockey, healthcare, and education. Designed to address the complexities of working with large, real-world databases, BIRD emphasizes the importance of understanding database values, incorporating external knowledge for accurate SQL query generation, and focusing on SQL execution efficiency. BIRD not only presents new challenges for developing more effective text-to-SQL models but also sets a new benchmark for evaluating model performance in real-world scenarios. It highlights the gap between current machine learning models and human-level SQL query formulation. Figure~\ref{Fig_sample} presents one example of data sample from BIRD benchmark.

\subsection{Experiment Setup}

We utilized a range of large language models (LLMs), such as T5 and GPT variants, each configured in unique setups to investigate various enhancement strategies, including the integration of external knowledge, prompt engineering, and feedback mechanisms. Each LLM was fine-tuned with a dataset tailored specifically to Text-to-SQL tasks, ensuring familiarity with the relevant vocabulary and structure. The models were tested under controlled conditions to precisely measure their performance in generating SQL statements.

%Other considerations for the experiments include:
%
%\begin{itemize}
%\item \textbf{Computational Resources:} The experiments were conducted on high-performance NVIDIA V100 workstations to handle the significant computation demands of LLMs.
%\item \textbf{Hyper-parameter Tuning:} Careful tuning of model hyper-parameters was carried out to optimize performance, exploring variations in learning rates, batch sizes, and the number of training epochs.
%\item \textbf{Version Control:} Each experimental setup, including model versions and configurations, was meticulously documented to ensure reproducibility and facilitate comparative analysis.
%\end{itemize}

\subsection{Evaluation Metrics}

To assess the performance of LLMs in generating SQL statements, two primary metrics were employed~\cite{li2024can}:

\begin{itemize}
    \item {Execution Accuracy (EX):} This metric quantifies the correctness of the generated SQL statements by comparing them to a set of predefined ground truths. A higher EX score indicates a higher similarity between the generated SQL statements and the ground truth, reflecting the model's understanding and query generation capabilities.
    
    \item {Valid Efficiency Score (VES):} VES evaluates the efficiency of the SQL statements generated by the models, taking into account both the accuracy of the SQL statement and its execution time against the database. A higher VES signifies not only correct but also optimally efficient generation of SQL statements, which is crucial for real-time applications of autonomous systems.

\end{itemize}

These metrics offer a comprehensive view of the models' performance, highlighting not just the accuracy but also the efficiency of the generated SQL statements, two critical aspects in the deployment of autonomous systems across various sectors.

\subsection{Result and Discussion}

Table~\ref{tab_result} presents a performance comparison between baseline models (T5 and GPT-3) and proposed methods based on GPT-3.5 and GPT-4, using EX and VES scores. Overall, incorporating knowledge (KG) significantly enhances the performance of LLM-based Text-to-SQL models. Additionally, when comparing the performance of LLMs using human-designed step-by-step prompts (HDSP) and feedback mechanisms, the feedback mechanism proves to be more beneficial. Furthermore, combining HDSP with the feedback mechanism allows GPT-4 to achieve the highest EX values, indicating that feedback effectively enhances the prompts within the HDSP framework shown in Figure~\ref{Fig_framework}. However, this combination reduces the VES values due to increased complexity in executing SQL statements, particularly with the feedback mechanism.

\begin{table}[h]
\centering
\caption{Comparing baselines with proposed methods by EX and VES scores, where KG and HDSP denote knowledge and human-designed step-by-step prompts, respectively. }
\begin{tabular}{l|l|c|c}
\hline
\multicolumn{2}{c|}{\textbf{LLMs}} & \textbf{EX} & \textbf{VES} \\
\hline

\multirow{6}{*}{T5} &T5-Base	& 	6.32 		&	7.78\\

&T5-Large  				& 	9.71		&	9.90\\
&T5-3B 					& 	10.37	&	13.62\\
&T5-Base + KG 			& 	11.54	&	12.90\\
&T5-Large + KG 			& 	19.75	&	22.74\\
&T5-3B + KG				& 	23.34	&	25.57\\
\hline
\multirow{2}{*}{GPT-3} & GPT-3  	& 24.05 & 27.97 \\
& GPT-3 + KG 					& 37.22 & 43.81 \\
\hline
\multirow{7}{*}{GPT-3.5} &GPT-3.5  & 26.01 & 29.39 \\
&GPT-3.5 + KG 		&  39.48		& 41.39 \\
%&GPT-3.5 + HDSP & 29.57&29.91 \\
%&GPT-3.5 + Feedback 	& 26.01 	& 29.16 \\
&GPT 3.5 + KG + HDSP 	&38.20	&39.25 \\
&GPT-3.5 + KG + Feedback 	& 40.48 	& 42.68 \\
&GPT-3.5 + KG + Feedback + HDSP 	&41.16  	& 41.37 \\
\hline
\multirow{7}{*}{GPT-4} &GPT-4  	&30.90  	&  34.60\\
&GPT-4 + KG 					& 46.35 	& \textbf{49.77} \\
%&GPT-4 + HDSP 				& 32.41	&35.68 \\
%&GPT-4 + Feedback 			& 30.28 & 31.28 \\ 
&GPT-4 + KG + HDSP 			& 47.06 	&47.26 \\
&GPT-4 + KG + Feedback 		& 47.86 & 47.35 \\
&GPT-4 + KG + Feedback + HDSP 	& \textbf{49.87} & 47.32 \\
\hline
\end{tabular}
\label{tab_result}
\end{table}

\begin{figure}[htbp]
    \centering
    \begin{subfigure}{0.5\textwidth}
        \centering
        \includegraphics[width=\textwidth]{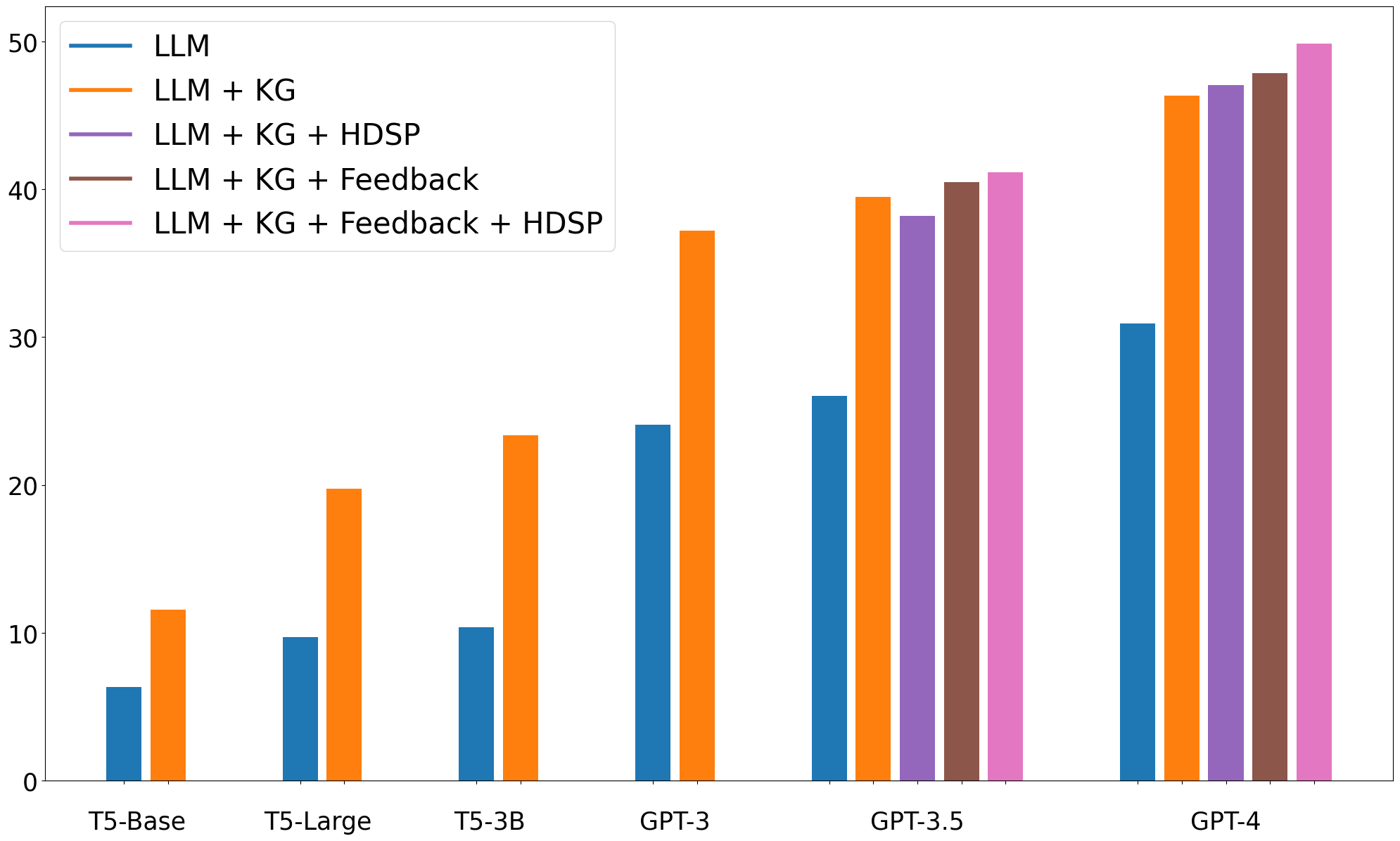}
        \caption{EX scores across different LLMs configurations.}
        \label{fig:overall_ex}
    \end{subfigure}
    \hfill
    \begin{subfigure}{0.5\textwidth}
        \centering
        \includegraphics[width=\textwidth]{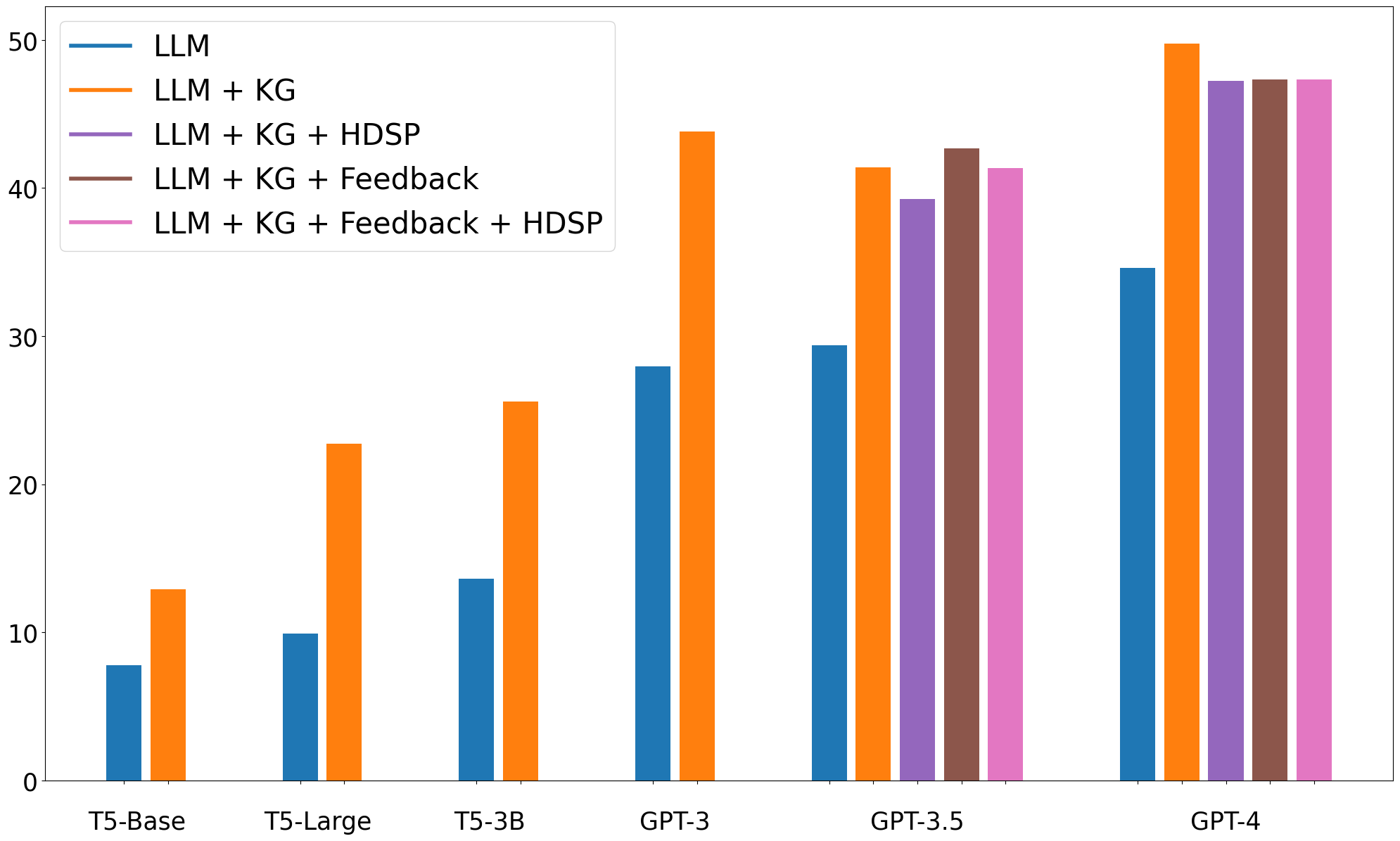}
        \caption{VES scores across different LLMs configurations.}
        \label{fig:ves_scores}
    \end{subfigure}
    \caption{A bar chart provides a clear visualization of the performance of different LLMs configurations.}
    \label{fig:comparison}
\end{figure}

Furthermore, Figure~\ref{fig:comparison} provides a clear visualization of the performance across different LLM configurations, making it easier to recognize performance trends. For EX scores, GPT-4 outperformed other LLMs when combined with HDSP and feedback. However, for VES scores, it was observed that the use of feedback and HDSP actually reduced performance.

Since GPT models outperform T5, we focus on GPT-based Text-to-SQL performance across different sample types. Tables~\ref{tab_ex} and~\ref{tab_ves} show the performance of GPT models on three sample types as predefined in the BIRD datasets. Incorporating knowledge (KG) enhances performance across all sample types, indicating that additional information about the database, such as schema and column details, aids in generating SQL statements. The feedback mechanism is particularly beneficial for improving performance on simple samples compared to moderate and difficult ones. Additionally, combining the feedback mechanism with human-designed step-by-step prompts (HDSP) improves performance for both simple and difficult samples with GPT-4. This validates that enhanced prompts with feedback are more effective in improving Text-to-SQL performance.

\begin{table}[h]
\centering
\caption{EX scores across different GPT model configurations.}
\begin{tabular}{l|l|c|c|c}
\hline
\multicolumn{2}{c|}{\textbf{LLMs}} & \textbf{Simple} & \textbf{Moderate} & \textbf{Difficult}\\
\hline
\multirow{2}{*}{GPT-3} & GPT-3   	& 31.08 	& 13.29  & 12.08\\
& GPT-3 + KG 					&45.44   	& 26.14  & 19.01\\
\hline
\multirow{7}{*}{GPT-3.5} &GPT-3.5 & 34.70 & 13.33 &11.11 \\
&GPT-3.5 + KG 		&  49.73 &   28.39 &18.06\\
%&GPT-3.5 + HDSP   		& 35.57  	&16.13   & 13.19\\
%&GPT-3.5 + Feedback 	& 34.81  	& 13.33  &10.42\\
&GPT 3.5 + KG + HDSP 	& 47.03  	& 26.24  &20.14\\
&GPT-3.5 + KG + Feedback 	& 50.05 	& 28.17   & 18.75\\
&GPT-3.5 + KG + Feedback + HDSP 	& 48.68 & 32.18 & 20.83  \\
\hline
\multirow{7}{*}{GPT-4} &GPT-4  	& 36.96 	& 24.64 &12.32 \\
&GPT-4 + KG 					&55.44  	&36.66  & 18.48\\
%&GPT-4 + HDSP   				& 38.48 	& 26.46 &12.68 \\
%&GPT-4 + Feedback 	& 40.32& 13.98& 15.28\\
&GPT-4 + KG + HDSP 	& 	42.78 &15.98& 17.28	\\
&GPT-4 + KG + Feedback & 56.86 & 24.87&  33.64\\
&GPT-4 + KG + Feedback + HDSP & 62.86 & 26.83&  35.68	\\
\hline
\end{tabular}
\label{tab_ex}
\end{table}

\begin{table}[h]
\centering
\caption{VES scores across different GPT Model configurations}
\begin{tabular}{l|l|c|c|c}
\hline
\multicolumn{2}{c|}{\textbf{LLMs}} & \textbf{Simple} & \textbf{Moderate} & \textbf{Difficult}\\
\hline
\multirow{2}{*}{GPT-3} & GPT-3 & 36.20 & 15.43  &14.42\\
& GPT-3 + KG 					& 54.71  	& 28.16 &22.80  \\
\hline
\multirow{7}{*}{GPT-3.5} &GPT-3.5   &40.05 &13.95& 10.76\\
&GPT-3.5 + KG 		& 51.83& 28.91&16.72\\
%&GPT-3.5 + HDSP   		& 37.46 	& 20.90 &12.07 \\
%&GPT-3.5 + Feedback 	&  39.80 	&13.93   &9.98\\
&GPT 3.5 + KG + HDSP 	&49.14  	& 27.43 &15.84\\
&GPT-3.5 + KG + Feedback &51.98& 28.83& 17.08	\\
&GPT-3.5 + KG + Feedback + HDSP & 48.65 & 32.28&20.71	 \\
\hline
\multirow{7}{*}{GPT-4} &GPT-4  	& 43.31 	& 24.17 &13.96 \\
&GPT-4 + KG 					& 60.31 	&34.76  &20.08 \\
%&GPT-4 + HDSP   				& 44.68 	&24.92  &14.40 \\
%&GPT-4 + Feedback 			& 41.07  	& 14.20  & 15.09 \\
&GPT-4 + KG + HDSP 		& 61.71 	&25.63  & 31.55	 \\
&GPT-4 + KG + Feedback 		&  60.43 	& 24.10  & 31.68\\
&GPT-4 + KG + Feedback + HDSP 	& 61.98  	& 23.82  & 31.87 \\
\hline
\end{tabular}
\label{tab_ves}
\end{table}

\section{Related Work}
\label{sec4}
% related work

Text-to-SQLs, which uses natural language queries to generate SQL statements, particularly through LLMs like GPT, is a rapidly evolving field. It aims to simplify the creation of SQL statements, enabling users to retrieve information using natural language. However, current performance of LLMs in Text-to-SQLs does not yet meet application requirements compared to human performance~\cite{li2024can}. To advance Text-to-SQL, the BIRD benchmark~\cite{li2024can}  has been proposed to assess the readiness of LLMs as database interfaces through text-to-SQL tasks. BIRD features a comprehensive dataset across multiple domains to challenge models with real-world database values and the need for external knowledge in SQL statement generation. 

Based on BIRD, several significant works have been completed to promote Text-to-SQLs. Sun~\textit{et. al} introduced the SQL-PaLM framework~\cite{sun2023sql} through using few-shot prompting and instruction fine-tuning. In few-shot prompting, it explores the effectiveness of consistency decoding combined with execution-based error filtering. Instruction fine-tuning delves into understanding the critical paradigms that influence the performance of fine-tuned LLMs. Additionally, it proposes a test-time selection method to further refine accuracy by integrating SQL outputs from multiple paradigms with execution feedback as guidance. Pourreza~\textit{et. al} proposed DIN-SQL~\cite{pourreza2024din}, which decomposes the Text-to-SQL task into four sub-steps: schema linking, complexity classification, SQL prediction, and self-correction. It also leverages few-shot prompting techniques. Experiments show that DIN-SQL not only narrows the performance gap between fine-tuned models and prompting approaches but also achieves state-of-the-art accuracy on the Spider and BIRD benchmarks. Li~\textit{et. al} proposed CodeS~\cite{li2024codes}, a series of pre-trained language models designed specifically for generating SQL statements from natural language inputs. Ranging from 1B to 15B parameters, CodeS models aimed to overcome the limitations of closed-source LLMs by being fully open-source and using an incremental pre-training approach on a curated SQL-centric corpus. CodeS enhances Text-to-SQL performance with comprehensive database prompts and a bi-directional data augmentation technique to facilitate adaptation to new domains. Wang~\textit{et. al} proposed a multi-agent system~\cite{wang2023mac} to enhance SQL statement generation from natural language queries. It consists of three agents: the Decomposer, which breaks down complex questions; the Selector, which filters relevant database schema elements; and the Refiner, which corrects SQL errors. The framework leverages chain-of-thought reasoning and few-shot learning, significantly improving accuracy and robustness in SQL statement generation. 

Although previous research has advanced LLM-based Text-to-SQL systems, these approaches often require complex data preprocessing or data augmentation techniques. In this study, we aim to enhance LLM-based Text-to-SQL by using straightforward feedback mechanisms and human-designed step-by-step prompts.

\section{Conclusion and Future Work}
\label{sec5}
% conclusion and future work

Recent advancements in LLM-based Text-to-SQL systems have improved the user experience for non-experts in retrieving information from relational databases using natural language queries. However, current state-of-the-art (SOTA) LLM-based Text-to-SQL models often require significant effort to develop auxiliary tools like SQL classifiers for optimal performance. This paper proposes a novel approach that enhances LLM-based Text-to-SQL performance using feedback based solely on SQL Quality Measurement. The proposed method undergoes comprehensive validation on the BIRD benchmark, assessing EX and VES across various Text-to-SQL difficulty levels. Experimental results reveal competitive performance in both EX and VES compared to SOTA models like GPT-4 and T5. Future work will involve validating the proposed method with more advanced LLMs and improving VES performance by reducing the complexity of feedback mechanisms.

%

\section*{Acknowledgment}
This research work is supported by NASA under award number 80NSSC22KM0052. Any opinions, findings, and conclusions or recommendations expressed in this material are those of the author(s) and do not necessarily reflect the views of NASA.



\bibliographystyle{IEEEtran}
\bibliography{References}
%\begin{thebibliography}{1}


%\end{thebibliography}

\end{document}